\documentclass[fp,twocolumn]{jpsj3}
\usepackage{txfonts,graphicx,bm,color,ulem}
\usepackage{overpic}
\def\tblu{}
\newcommand{\bmq}{\bm{q} }
\newcommand{\bmp}{\bm{p} }
\newcommand{\gfac}{\gamma}

\title{Nuclear Magnetic Relaxation Time near the  Compensation Temperature in a Ferrimagnetic Insulator}

\author{Michiyasu Mori}
\inst{Advanced Science Research Center, Japan Atomic Energy Agency, 
Tokai, Ibaraki 117-1195, Japan} 
\abst{
The nuclear magnetic relaxation time $T_1$ in a ferrimagnetic insulators is 
calculated 
within the mean-field approximation for the magnetic exchange interactions and the Raman process involving the hyperfine interaction. 
We find that the value of 1/$T_1$ on one type of site increases rapidly near the compensation 
temperature $T_0$, whereas that on the other type of site does not increase up to Curie 
temperature $T_c$. 
This is due to the fact that the soft-magnon bandwidth becomes 
comparable to $T_0$. 
An increase in 1/$T_1$ below $T_c$ is found also in another type 
ferrimagnet, which shows a hump structure in the temperature dependence of 
magnetization instead of compensation. 
Also in that case, we find the rapid increase in 1/$T_1$ below 
$T_c$, even though the magnetization does not show compensation. 
The coexistence of soft and hard magnons leads to these remarkable 
properties of ferrimagnets. 
}
\date{\today}

\begin{document}
\maketitle
\section{Introduction}
A {\it ferri-}magnet is a kind of {\it ferro-}magnet, and it was theoretically predicted by N\'eel\cite{neel48,neel63,neel71}. 
Soon afterward, the magnetization compensation was observed in the LiFeCr spinel ferrite, 
for which the magnetization becomes zero at magnetization-compensation temperature $T_{\rm M}$ far below the Curie temperature $T_{\rm c}$\cite{gorter53}. 
Such a ferrimagnet, called an N-type ferrimagnet, also has been found in rare-earth iron garnets (RIGs)\cite{neel63,pauthenet54,bertaut56,geller57acta,geller57,pauthenet58,geller63,geller65}.  The RIGs have been studied by many authors in order to apply their magnetization-compensation properites to magneto-optical memories\cite{chang65,chow68,nelson68}.

The dynamical aspects of ferrimagnetism were initially studied using the electron spin resonance (ESR)\cite{vanwieringen53,kaplan53,wangsness53,tsuya54,wangsness54,wangsness55,mcgire55,geshwind59,vanvleck61}. 
The ferrimagnetic resonance (FIR) differs from the ferromagnetic resonance (FMR) 
in having two branches. 
One gives the usual FMR, while the other, called the exchange frequency, is located higher in energy \cite{wangsness55}. 
It was difficult to measure the exchange frequency when it was first discovered, since its wavelength is of the order of a tenth of a millimeter. 
However, a singular behavior of the gyromagnetic ratio was observed around the angular momentum compensation temperature $T_A$ in a LiFeCr spinel ferrite\cite{vanwieringen53,mcgire55}. 
On the lower branch, the  effective gyromagnetic ratio becomes small around $T_{\rm M}$ and then increases rapidly around $T_{\rm A}$\cite{vanwieringen53,kaplan53,wangsness53,wangsness54,tsuya54,wangsness55,mcgire55,geshwind59,vanvleck61}. 
The $g$-value of the upper branch becomes small in a measurable range around $T_A$\cite{mcgire55}. 
The  magnetization is the  product of the Lande $g$-factor and a total angular momentum. 
In general, hence, $T_{\rm M}$ is different from $T_{\rm A}$, when the  orbital angular momentum is involved. 
In contrast to the magnetization, the dynamics of a ferrimagnet become singular around $T_A$. 

Because the magnetization couples to a magnetic field, while the  total angular momentum itself does not, it can be difficult to measure $T_A$ directly using conventional methods. Recently, however, Imai et al. have successfully observed $T_A$ using the Barnett effect\cite{imai18,imai19,imai20}. 
In a rotating frame, the  rotation frequency couples to the  angular momentum instead of to the magnetization, without any coupling constant. 
By spin-rotation coupling, a magnetization is induced through the  angular momentum by mechanical rotation. 
This was originally studied by Barnett\cite{barnet15}, and it is now used to determine $T_A$ in a RIG\cite{imai18,imai19}. 
It has been reported that around $T_A$ the  magnetization reverse rapidly and that domain walls move fast\cite{jiang06,stanciu07,kjkim17,skkim19}. 
Those properties, which are advantageous for magnetic memories, are attributed to angular-momentum compensation.  

Nuclear magnetic resonance (NMR) also is a powerful tool for studying the magnetism of a broad range of materials. 
Magnetic excitations can be characterized by the nuclear magnetic relaxation time $T_1$, 
which originates in the hyperfine interaction between electron and nucleus. 
For magnetic insulators, however, the origin of $T_1$ is not so obvious. 
If the system is isotropic and the nuclear and electron quantization axes are identical, the relaxation cannot be obtained within the linearized spin-wave approximation. 
Misalignment of the quantization axes--and/or the dipole-dipole 
interactions between an electronic and a nuclear spins--induces 
relaxation through the Raman process\cite{moriya56,mitchell57,beeman68}. 
Interactions among magnons are also the source of relaxation, e.g., through the three-magnon process\cite{moriya56,mitchell57,beeman68}. 
Those processes can be studied for both ferromagnetic and antiferromagnetic insulators. 
Recently, Imai et al. have reported an enhancement of the NMR signal around $T_{\rm A}$, which is closely related to domain wall motion\cite{imai20}.  
In contrast to ESR, NMR provides a site-selective measurement of magnetism. 
It is therefore interesting to study the dynamical aspects of magnetism site-by-site in a ferrimagnet.  
In addition, a consistent understanding of ferrimagnetism among experimental methods--NMR, ESR, and neutron scattering--will be useful. 

In this paper, we study the nuclear magnetic relaxation time in ferrimagnets. 
Section 2 explains the model Hamiltonian and the approximation used. 
The nuclear magnetic relaxation time due to the Raman process is given in Sec. 3. 
Additional changes due to orbital angular momentum are briefly discussed in Sec. 4. 
Below, Bohr magneton $\mu_B$ and Planck constant $\hbar=h/2\pi$ are set equal to 1 for brevity. 

\section{Formalism: Magnons in Ferrimagnet}
\begin{figure}[htbp]
	\centering
	\includegraphics[width=0.5\textwidth]{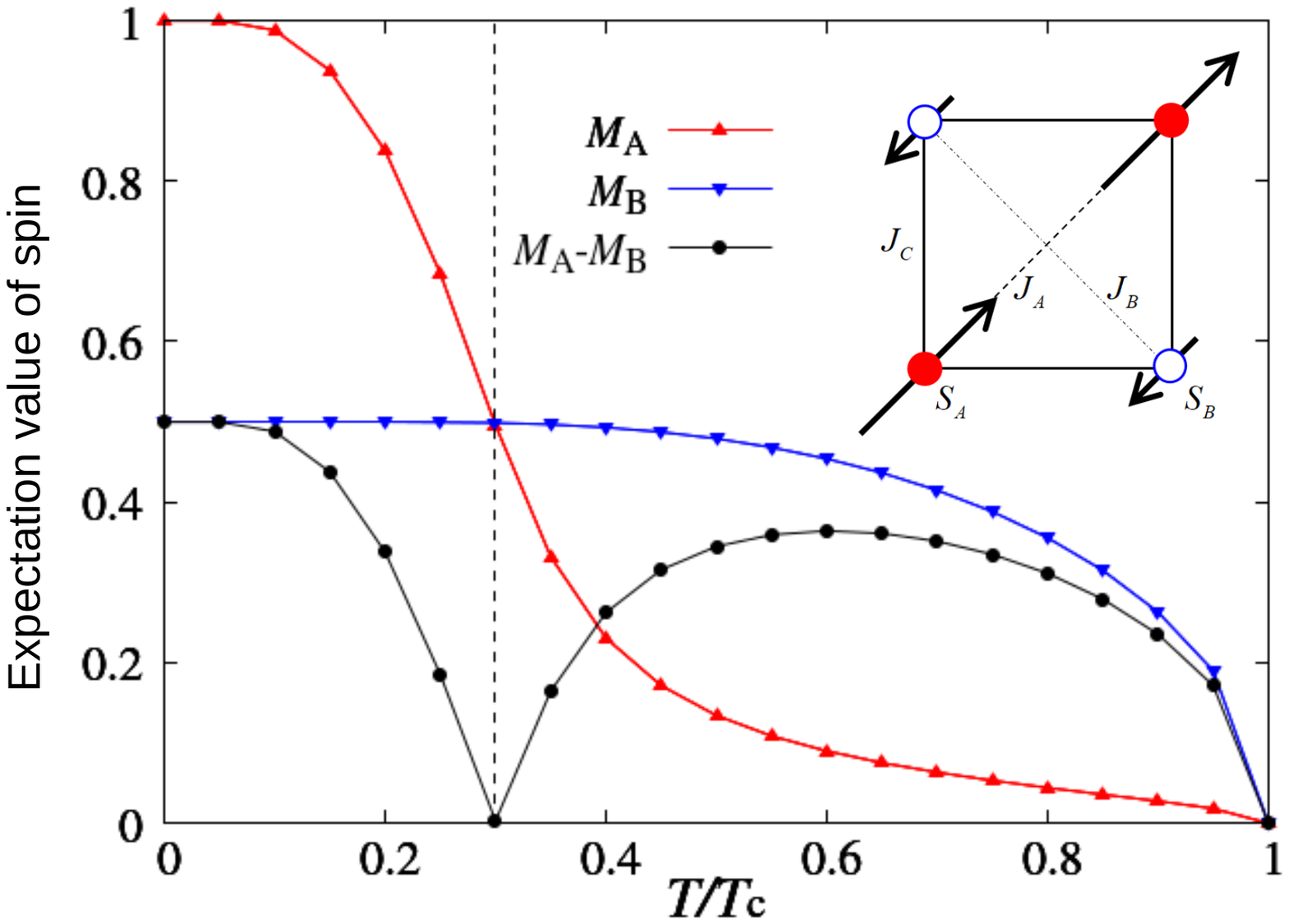}
	\caption{(Color online) The mean-field solution for $J_A$=0.1, $J_B$=1.0, $J_C$=0.05, 
		$S_A$=1, and $S_B$=1/2. The inset shows the lattice structure, which is three 
		dimensional. The red 
		(upward) and blue (downward) triangles denote $M_A$ and $M_B$, 
		respectively. The black circles are 
		the sum of two expectation values, $M_A-M_B$. The broken line indicates 
		$T_0$.}
	\label{mfsol}
\end{figure}

We will focus on a ferrimagnetic "insulator," which is simply called 
a "ferrimagnet" below. 
The magnetic exchange interaction due to the Pauli principle and to the Coulomb 
interaction between electrons is the source of magnetism in a ferrimagnet.
Two sub-lattices with different spin magnitudes $S_A \neq S_B$ comprise the  
simplest model. The Hamiltonian is given by 
\begin{align}
H
&=
-J_A\sum_{\left\langle {i,i'} \right\rangle } \vec{S}_i \cdot \vec{S}_{i'}
-J_B\sum_{\left\langle {j,j'} \right\rangle } \vec{S}_j \cdot \vec{S}_{j'}
+J_C\sum_{\left\langle {i,j } \right\rangle } \vec{S}_i \cdot \vec{S}_j,\label{heisenberg}
\end{align}
with the spin operators $\vec{S}_i$ ($\vec{S}_j$) on site $i$ $\in$ $A$-sites 
($j$ $\in$ $B$-sites). 
The angular brackets $\langle\cdots\rangle$ denotes nearest neighbor sites. 
The magnitudes of the magnetic exchange interactions $J_A$, $J_B$, and $J_C$ are assumed to be positive for brevity.
First, we do not consider the  orbital angular momentum $\vec{L}_i$.
Hence, 
there is only one 
compensation temperature $T_0$; i.e., $T_M$ $=$ $T_A$ $\equiv$ $T_0$.   
What is changed by $\vec{L}_i$ will be discussed 
in the last section. 
At $T_0$, the expectation values $\langle S_A^z \rangle$ $\equiv$ $M_A >0$ and 
$\langle S_B^z 
\rangle$ $\equiv$ $-M_B <0$ satisfy $M_A-M_B$ $=$ 0, where the bracket 
denotes 
the thermal average. See also Appendix\ref{B}. 
It is known that there are some possible cases of compensation.
In the case considered above, both sub-lattices have the same number of sites 
in a unit cell, as shown in the inset of Fig. \ref{mfsol}.  
Another case has $M_A n_A -M_B n_B$ = 0, for which the number $n_A$ of spins on sub-lattice $A$ is different from the number $n_B$ on sub-lattice $B$. 
As shown in Appendix\ref{A}, those lattice structures have characteristic features 
in common. 
Hence, we consider the simplest case, shown in Fig. 
\ref{mfsol} below. 

Compensation occurs at a finite temperature. 
To include the temperature dependences of $M_A$ and $M_B$, 
we adopt the mean-field approximation and use the linearized spin-wave approximation 
around the mean-field solution. 
This is equivalent to Tyablikov decoupling in the Green's function method 
and is a kind of random-phase approximation~\cite{tyablikov59,oguchi63}.  
The mean-field solution for $J_A$=0.1, $J_B$=1.0, $J_C$=0.05, $S_A$=1, and 
$S_B$=1/2, is shown in Fig. \ref{mfsol}, where $T_0/T_c$ $\sim$ 0.3 and the 
Curie temperature is $T_c$ $\sim$ 3.0.

The Holstein-Primakoff (HP) bosons (magnons)--for which the creation and annihilation operators are $a_i^\dag, a_i$ on 
$A$-sublattice  
and 
$b_j^\dagger, b_j$ on $B$-sublattice--are given by,
$S_i^-$ $\sim$ $\sqrt{2M_A} a_i^\dag$, 
$S_i^+$ $\sim $$\sqrt {2M_A}a_i$, 
$S_i^z$ $=$ $M_A$ $-$ $a_i^\dag {a_i}$, 
$S_j^+$ $\sim$ $\sqrt{2M_B} b_j^\dag $, 
$S_j^-$ $\sim $$\sqrt{2M_B}b_j$, 
$S_j^z$ $=$ $b_j^\dag {b_j}$ $-$ $M_B$. 
Below, the spins are assumed to be ordered in the $z$-direction.
By the linearized approximation, the action of the magnons is given 
by\cite{mori17}
\begin{align}
	S&=\sum_{\bmq,i\omega_n} 
	\Phi^\dag\left[
	\left(
	\begin{array}{*{20}{c}}
		-i\omega_n&0\\
		0&i\omega_n
	\end{array}
	\right)
	+\left(
	\begin{array}{*{20}{c}}
		\varepsilon_{1\bmq}  & \varepsilon_{3\bmq}^*\\
		\varepsilon_{3\bmq}  & \varepsilon_{2\bmq}
	\end{array}
	\right)
	\right]\Phi,\label{action}\\
\Phi^\dag
	&\equiv
		\left(a_{\bmq}^\dagger(i\omega_n),b_{-\bmq}(-i\omega_n)\right),\\
\varepsilon_{1\bmq}
	&\equiv
		zJ_C M_B+z_A J_A M_A(1-\zeta_{A\bmq}), \label{eps1}\\ 
\varepsilon_{2\bmq}
	&\equiv
		zJ_C M_A+z_B J_B M_B(1-\zeta_{B\bmq}),\label{eps2}\\ 
\varepsilon_{3\bmq}
	&\equiv
		J_C\sqrt{M_A M_B}\sum_{\xi}e^{i\bmq\bm{\cdot\xi}},\label{xi}\\
\zeta_{A(B)\bmq}
	&\equiv
		\frac{1}{z_{A(B)}}\sum_\eta \cos(\bmq\bm{\cdot\eta}),\label{zeta}
\end{align}
We use the boson operators $a_{\bmq}(i\omega_n)$ and $b_{-\bmq}(-i\omega_n)$ with 
momentum $\bmq$$=$$(q_x,q_y,q_z)$ and Matsubara frequency $\omega_n$. 
The $\bmq$-summation is taken over the first Brillouin zone.
The magnon dispersion relation depends on the  connectivity of the sub-lattice, which gives the number of nearest-neighbor sites $z_A$ and $z_B$ on each sub-lattice, and the number $z$ of nearest-neighbor sites between the two sub-lattices. 
In Eqs. (\ref{xi}) and (\ref{zeta}), \bm{$\xi$} and \bm{$\eta$} mean the summation over the nearest-neighbor sites between the two sub-lattices and within each sublattice, respectively.  
From Eq. (\ref{action}), the magnon Green's functions $g_\nu(\bmq,i\omega_n)$ are given by
\begin{align}
g_A(\bmq,i\omega_n)
&\equiv \langle a_{\bmq}(i\omega_n)a_{\bmq}^\dag(i\omega_n)\rangle\nonumber\\
&= -\frac{i\omega_n + 
	\varepsilon_{2\bmq}}{\left(i\omega_n-E_{\alpha \bmq}\right)\left(i\omega_n+E_{\beta \bmq} 
	\right)},\label{ga}\\
g_B(\bmq,i\omega_n)
&\equiv \langle b_{\bmq}(i\omega_n)b_{\bmq}^\dag(i\omega_n)\rangle\nonumber\\
&=-\frac{i\omega_n + 
	\varepsilon_{1\bmq}}{\left(i\omega_n-E_{\beta \bmq}\right)\left(i\omega_n+E_{\alpha \bmq} 
	\right)},\label{gb}
\end{align}
and the magnon dispersion relations $E_{\alpha \bmq}$ and $E_{\beta \bmq}$ are given by
\begin{align}
E_{\alpha \bmq}
&=	\frac{1}{2}\left[\left(\varepsilon_{1\bmq}-\varepsilon_{2\bmq}\right)
+\sqrt{\left(\varepsilon_{1\bmq}+\varepsilon_{2\bmq} 
	\right)^2-4\left|\varepsilon_{3,\bmq}\right|^2}\right],\label{ea}\\
E_{\beta \bmq}
&=	\frac{1}{2}\left[-\left(\varepsilon_{1\bmq}-\varepsilon_{2\bmq}\right)
+\sqrt{\left(\varepsilon_{1\bmq}+\varepsilon_{2\bmq} 
	\right)^2-4\left|\varepsilon_{3,\bmq}\right|^2}\right].\label{eb}
\end{align}
The small-$Q$ approximation for 
$M_A$ $>$ $M_B$ leads to, 
\begin{align}
E_{\alpha \bmq}&\sim C Q^2,\\  
E_{\beta \bmq}&\sim 12J_C\left( {{M_A} - {M_B}} \right) + D Q^2,
\end{align}
where $C$ and $D$ are constants given in Appendix \ref{A} and $Q$ $\equiv$ $\sqrt{q_x^2+q_y^2+q_z^2}$. 
The mode $E_{\alpha q}$ is gapless, while $E_{\beta q}$ has an "optical 
gap," $E_g$ $\equiv$ $|E_{\alpha q=0}-E_{\beta q=0}|$ $=$ 
$12J_C\left( M_A - M_B \right)$, which disappears at $T_0$. 
The dispersion relations degenerate at the gamma point and 
increases linearly with $Q$, similar to an antiferromagnet. 
Away from the gamma point, on the other hand, the two dispersion relations 
deviate from each other:  
\begin{align}
E_{\alpha \bmq}&\sim C_1 Q + C_2 Q^2,\\
E_{\beta \bmq}&\sim C_1 Q -C_2 Q^2,
\end{align}
where $C_1$ and $C_2$ are constants given in Appendix \ref{A}. 
These $Q$-dependences are relevant to 
the temperature dependence of $T_1$ at low temperatures. 

\section{Results: Nuclear Magnetic Relaxation due to the Raman Process}
In this study, we consider the nuclear magnetic relaxation time $T_1$ 
originating from the contact interaction between a nucleus and an electron
\begin{align}
H_{\rm n-el} 
 &=\frac{1}{2}\sum_{\nu=A,B}\left[\gfac_\nu \sum\limits_i 
 f_{i\nu}\left( \vec{S_i}\cdot\vec{I_i} \right)\right],\label{cont}
\end{align}
with the $g$-factor on $\nu$-sites being given by $\gfac_\nu$ and the nuclear spin $\vec{I_i}$ on $i$-site. 
If the system is isotropic and the quantization axes of the nucleus and electron 
are identical, we cannot obtain relaxation within the linearized 
approximation. 
Misalignment of the quantization axes--and/or the dipole-dipole 
interactions between electronic and nuclear spins--will induce relaxation due to the Raman process\cite{moriya56,mitchell57,beeman68}. 
Interactions among magnons also cause relaxation, e.g., through the three 
magnon-process\cite{moriya56,mitchell57,beeman68}. 
Below, we focus on the Raman process induced by misalignment. 
This is sufficient to enable us to find some of the characteristics of $T_1$ near $T_0$.
The critical exponent of $T_1$ is beyond the scope of this 
study and will be discussed elsewhere. 
When the quantization axis of nucleus deviates by an angle $\theta$ from 
that of the electron, 
Eq. (\ref{cont}) reduces to the following component
\begin{align}
H^z_{\rm n - el} 
	&= \frac{1}{2} \sum_{\nu=A,B} 
		\left[\gfac_\nu\sum\limits_i \sin\theta f_{i\nu} S_i^z 
		\left(I_i^++I_i^-\right)\right], \label{raman}
\end{align}
which are relevant for calculating $T_1$.  
Assuming that $\gfac_A$ $=$ $\gfac_B$ $\equiv$ $\gfac$  and the form factors $f_{iA}$ are constant, i.e., $f_{iA}$ $=$ 
$f_{iB}$ $\equiv$ $f$, the $T_1$ on site $\nu$ $=A, B$ is given by
\begin{align}
\frac{1}{T_{1\nu}} 
&= F \sum\limits_{\bmq} C_\nu(\bmq,\omega_0),\\
C_\nu(\bmq,\omega_0)
	&=\int {dt\;{e^{i\omega_0 t}}} \left\langle {S_{\nu \bmq}^z(t)S_{\nu,-\bmq}^z(0) 
	+ S_{\nu,-\bmq}^z(t)S_{\nu \bmq}^z(0)} \right\rangle,\nonumber\\
\end{align}
where $\langle \cdot\cdot\cdot \rangle$ means the thermal average. 
The nuclear magnetic resonance energy is denoted by $\omega_0$, and 
$F = (\gfac f \sin\theta /2)^2$. 
Using Eqs. (\ref{ga}) and (\ref{gb}), 
the spin-spin correlation function $C_\nu(\bmq,\omega_0)$ is given by
\begin{align}
C_\nu(\bmq,\omega_0)
&= \frac{2}{1-{\rm e}^{\omega_0/k_BT}}{\rm Im} \Pi_\nu^R(\bmq,\omega_0),\\
\Pi_\nu(\bmq,i\omega_0)
&=k_BT\sum_{\bmp,n}g_\nu(\bmp+\bmq,i\omega_n+i\omega_0)g_\nu(\bmp,i\omega_n),
\end{align}
where $T$ is temperature, $k_B$ is the Boltzmann's constant, and momentum $\bmp$ = ($p_x$,$p_y$,$p_z$) in the first Brillouin zone.  
The retarded function of $\Pi_\nu(\bmq,i\omega_0)$ is denoted by 
$\Pi_\nu^R(\bmq,\omega_0)$. 
When $\omega_0$ is much smaller than $k_BT$, the nuclear magnetic relaxation time $T_{1 \nu}$ on site $\nu$ is given by
\begin{align}
\frac{1}{T_{1\nu}} 
	&= 2 F \sum_{\bmq} 
		\lim_{\omega_0\rightarrow 0}\frac{k_BT}{\omega_0}{\rm Im} \Pi_\nu^R(\bmq,\omega_0),\\
	&=2\pi F 
		\sum_{\bmp,\bmq} \biggl\{ n_B(E_{\nu \bmp})\left[n_B(E_{\nu \bmp})+1 \right]
			\frac{\alpha_{\bmp}\alpha_{\bmq}}{\Delta_{\bmp}\Delta_{\bmq}}\delta \left( E_{\nu {\bmp}} - 
			E_{\nu \bmq} \right) \nonumber\\
	& +n_B(E_{\mu \bmp})\left[n_B(E_{\mu \bmp})+1 \right]
			\frac{\beta_{\bmp}\beta_{\bmq}}{\Delta_{\bmp}\Delta_{\bmq}}\delta \left( E_{\mu \bmp} - 
			E_{\mu \bmq} \right)\biggr\},\label{t1inv}\\
\frac{\alpha_{\bmp}}{\Delta_{\bmp}} 
	&=\frac{1}{2}\left(\frac{\varepsilon_{1\bmp} + 
	\varepsilon_{2\bmp}}{\Delta_{\bmp}}+1\right),\\
\frac{\beta_{\bmp} }{\Delta_{\bmp}}
	&=\frac{1}{2}\left(\frac{\varepsilon_{1\bmp} + 
	\varepsilon_{2\bmp}}{\Delta_{\bmp}}-1\right),\\
\Delta_{\bmp}&=\sqrt {{{\left( {{\varepsilon_{1\bmp}} + {\varepsilon_{2\bmp}}} \right)}^2} 
- 4{{\left| {{\varepsilon_{3\bmp}}} \right|}^2}},
\end{align}
Here $\mu\neq\nu$, i.e., $\mu$=$\alpha$ for $\nu$=$\beta$ or $\mu$=$\beta$ for $\nu$=$\alpha$ and the Bose distribution function is denoted by $n_B(x)\equiv 1/[e^{x/(k_{B}T)}-1]$. 
Note that Eq. (\ref{t1inv}) can be checked by considering the ferro- and the 
antiferromagnet cases as discussed in Appendix \ref{C}.  

\begin{figure}[htbp]
	\centering
	\includegraphics[width=0.45\textwidth]{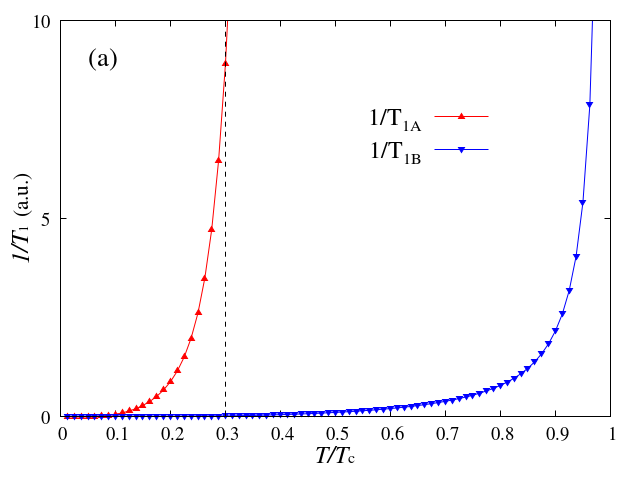}
	\includegraphics[width=0.45\textwidth]{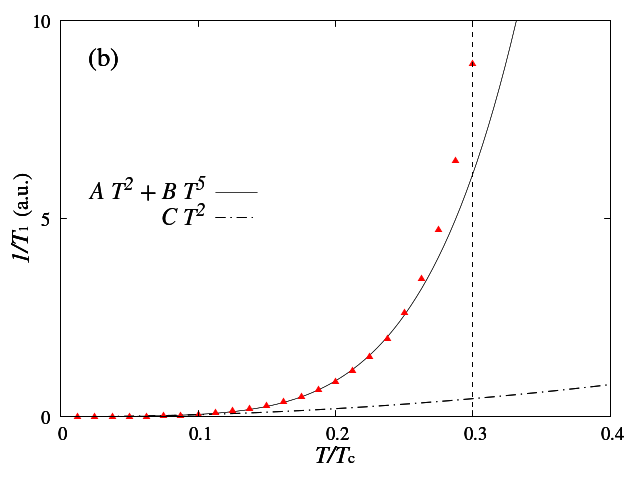}
	\caption{(Color online) (a) The $T$-dependence of $1/T_1$ on the A-sites and B-sites (1/$T_{1A}$ 
	and 1/$T_{1B}$) are plotted with 
	upward triangles (red) and downward triangles (blue), respectively. The 
	mean-field solution for $J_A$=0.1, $J_B$=1.0, $J_C$=0.05, $S_A$=1, and 
	$S_B$=1/2 was used. The broken line indicates $T_0$. See also Fig. 
	\ref{mfsol}. (b) At low temperatures, $T_{1A}$ is well fitted by $T^2$ 
	(the dashed-dotted line) similar to the ferromagnet (See also Appendix \ref{C}), 
	whereas it is deviated with increasing temperature as $A T^2$ $+$ $B T^5$ 
	(solid line). $A$ and $B$ are constants.}
\label{ferrinmr1}
\end{figure}

Using Eq. (\ref{t1inv}) and the mean-field solution shown in Fig. \ref{mfsol}, 
the $T$-dependence of 1/$T_{1\nu}$ can be calculated numerically as shown in Fig. 
\ref{ferrinmr1} (a).  
Note that $1/T_{1,A}$ increases rapidly around $T/T_c$ $\sim$ 0.3, 
which corresponds to $T_0$ indicated by the broken line in Fig. \ref{ferrinmr1}. 
This contrasts sharply with $1/T_{1B}$, which diverges just below $T_c$. 
As shown in Fig. \ref{ferrinmr1} (b), at low temperatures, $T_{1A}$ is well 
fitted by $T^2$, similar to the ferromagnet (See also Appendix~\ref{C}). 
With increasing temperature, on the other hand, it is fitted by  $A T^2$ $+$ $B 
T^5$ with constants $A$ and $B$. This is similar to the behavior of a ferromagnet, 
except for the fact that $1/T_{1A}$ increases around $T_0$ instead of $T_c$. 

\begin{figure}[htbp]
	\centering
	\includegraphics[width=0.4\textwidth]{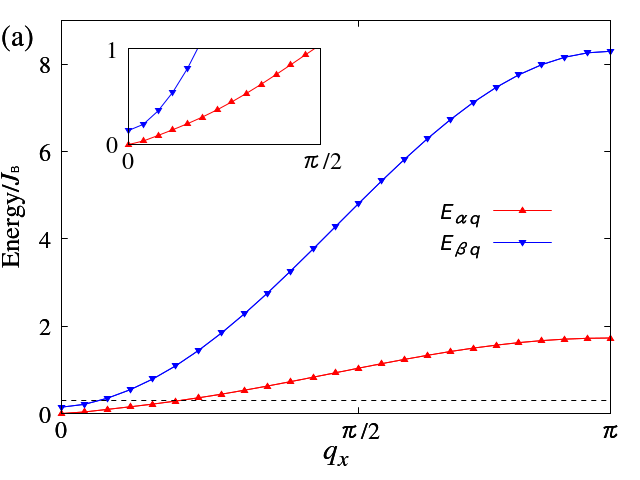}
	\includegraphics[width=0.4\textwidth]{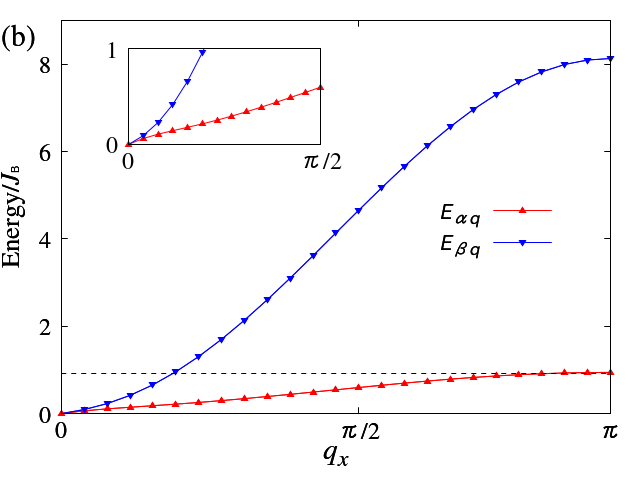}
	\caption{(Color online) The dispersion relations $E_{\alpha \bmq}$ and 
		$E_{\beta \bmq}$ are plotted by upward triangles (red) and downward triangles 
		(blue), respectively, for (a) $T/T_c$ $=$ 0.1 and (b) 
		$T/T_c$ $=$ 0.3, with $q_y$ $=$ $q_z$ $=$ 0. 
		In the pre-set lattice structure, at $\bmq=(\pi,\pi/2,0)$, $E_{\alpha \bmq}$ is a maximum. 
		This can be approximated by its value at $\bmq=(\pi,0,0)$ due to the small value of $J_C$. 
		The low-energy region is enlarged and plotted in the insets.}
\label{ferridisp} 
\end{figure}

To understand this behavior of 1/$T_1$ around $T_0$, $E_{\alpha\bmq}$ and 
$E_{\beta\bmq}$ are plotted in Fig. \ref{ferridisp} for (a) $T/T_c$ $=$ 0.1 and (b) 
 $T/T_c$ $=$ 0.3 with $q_y$ $=$ $q_z$ $=$ 0. 
At low temperatures, the $T$-dependence of 1/$T_1$ is determined by 
the $Q^2$-dependences of 
$E_{\alpha \bmq}$ and $E_{\beta \bmq}$ around $Q$ $\sim$ 0. See also the inset of Fig. 
\ref{ferridisp} (a). At $T$ $=$ $T_0$, $E_g$ becomes zero, as shown in 
the inset of Fig. \ref{ferridisp} (b), and both $E_{\alpha \bmq}$ and $E_{\beta \bmq}$ become proportional to 
$Q$ instead of $Q^2$ around $Q=0$.
Note that $k_B T/J_B$ is shown by the broken line in Figs. \ref{ferridisp} (a) and (b) as a measure of 
the  temperature. 
In the pre-set lattice structure, $E_{\alpha\bmq}$ is a maximum at $\bmq=(\pi,\pi/2,0)$. 
It can be approximated by the value at $\bmq=(\pi,0,0)$, since the  small value of $J_C$ $=$ 0.05. 
We then find that at $T$ $=$ $T_0$ the bandwidth of $E_{\alpha\bmq}$ becomes comparable to $k_B T$. 
This means that all states of $E_{\alpha\bmq}$ contribute to 1/$T_1$ through $n_B(x)$ in 
the 
first term in Eq. (\ref{t1inv}), where the first term is dominant. 
This is the origin of the rapid increase in 1/$T_{1A}$ at $T_0$. 
This is not accidental, due to the following points. 
The bandwidth of $E_{\alpha \bmq}$ can be very roughly estimated by
$\varepsilon_{1,\bmq=(\pi,0,0)}$$\sim$(6$J_C$+16$J_A$)$M_0$ $\sim$ 8$J_A$ = 0.8 with $M_A$ $=$ $M_B$ $\equiv$ 
$M_0$ $\sim$ 0.5. 
On the other hand, $T_0$ can be roughly estimated as 
$T_0$ $\sim$ $z_A$$J_A$$X_A$ = 0.8. Since $M_A$ is soft and decreases rapidly  
with $T$, $T_0$ is close to the Curie temperature of a system limited 
to the $A$-sublattice. 
Therefore, 1/$T_{1A}$ rapidly increases around $T_0$. 

\begin{figure}[htbp]
	\centering
	\includegraphics[width=0.45\textwidth]{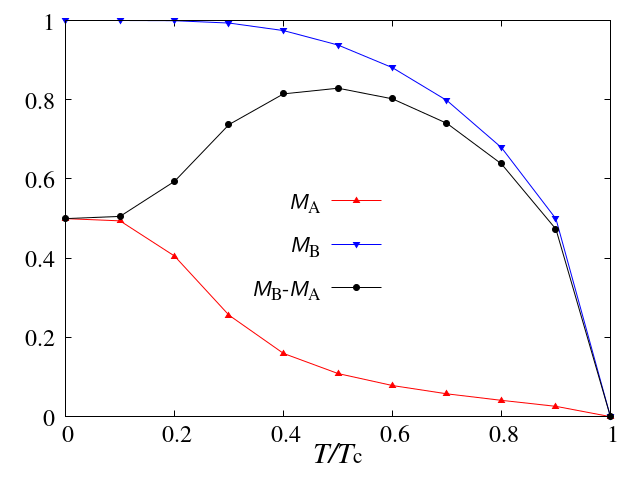}		
	\caption{(Color online) The mean-field solution for a $P$-type ferrimagnet with 
	$J_A$=0.5, $J_B$=1.0, $J_C$=0.2,  
			$S_A$=1/2, and $S_B$=1. }  
\label{ptype}
\end{figure}

\begin{figure}[htbp]
	\centering
	\includegraphics[width=0.45\textwidth]{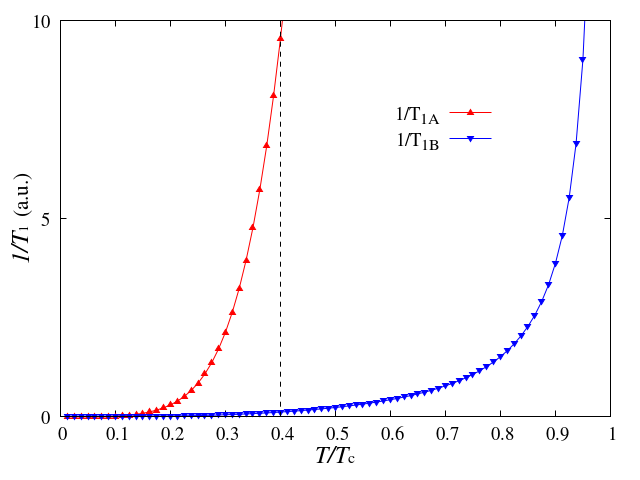}
	\caption{(Color online) The relaxation rate 1/$T_1$ in a $P$-type ferrimagnet. The $T$-dependences of $1/T_1$ on 
	the A-sites and B-sites (1/$T_{1A}$ and 1/$T_{1B}$) are plotted by 
	upward triangles (red) and downward triangles (blue), respectively. The 
	mean-field solution for $J_A$=0.5, $J_B$=1.0, $J_C$=0.2, $S_A$=1/2, and 
	$S_B$=1 was used. The broken line is near the top of the hump 
	structure.}
\label{ferrinmr7}
\end{figure}

\begin{figure}[htbp]
	\centering
	\includegraphics[width=0.4\textwidth]{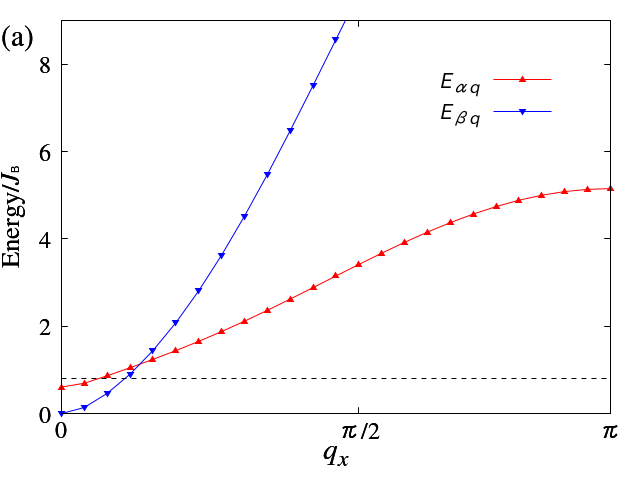}
	\includegraphics[width=0.4\textwidth]{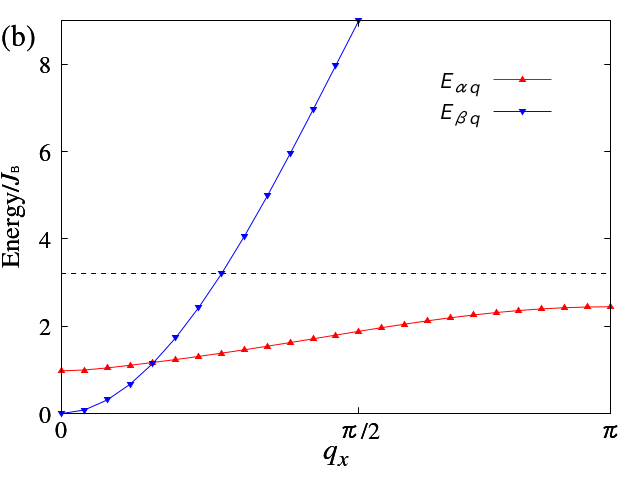}
	\caption{(Color online) The dispersion relations $E_{\alpha \bmq}$ and 
		$E_{\beta \bmq}$ are plotted by upward triangles (red) and downward triangles 
		(blue), respectively, for (a) $T/T_c$ $=$ 0.1 and (b) 
		$T/T_c$ $=$ 0.4, with $q_y$ $=$ $q_z$ $=$ 0. 
	The broken line indicates the corresponding temperature $k_B T/J_B$. }
\label{ferridisp3}
\end{figure}

So far, we have discussed an $N$-type ferrimagnet \cite{neel48,neel71}. 
Another type of ferrimagnet--called $P$-type--shows a hump in the  temperature 
dependence of the magnetization instead of compensation. 
Figure \ref{ptype} is calculated from Eq. 
(\ref{heisenberg}) for $J_A$=0.5, $J_B$=1.0, $J_C$=0.2, 
$S_A$=1/2, and $S_B$=1. 
The lattice structure is the same as the inset of Fig. \ref{mfsol}. 
In a $P$-type ferrimagnet, the magnetization does not show any singular behavior, such as 
compensation, 
although it is composed of two different sublattices, i.e., with soft and hard dispersion relations $E_{\alpha\bmq}$ and $E_{\beta\bmq}$. 
The value of 1/$T_{1\nu}$ in a $P$-type ferrimagnet is plotted in Fig. \ref{ferrinmr7} 
in the same way as for the $N$-type. 
We find that 1/$T_{1A}$ rapidly increases around $T/T_c\sim$ 0.4 close 
to the top of the hump structure, while 1/$T_{1B}$ increases rapidly near $T_c$. 
It is now straightforward to understand this behavior, since the $A$-sublattice is 
soft and the $B$-sublattice is hard. 
This is clearly shown by the dispersion relation in Figs.~\ref{ferridisp3} for 
(a) $T/T_c$ $=$ 0.1 and (b) $T/T_c$ $=$ 0.4. 
In each panel, $k_B T/J_B$ is shown by the broken line as measure of 
the temperature. 
At $T/T_c$ $=$ 0.4, all of $E_{\alpha \bmq}$ contribute to 1/$T_{1A}$. 
Therefore, even in a $P$-type ferrimagnet without compensation, we find 
a rapid increase of 1/$T_1$. 

\section{Summary and Discussions}
We have studied 
$T_1$ in a ferrimagnetic 
insulator that is induced by the Raman process involving the hyperfine interaction. 
To calculate 1/$T_1$, we adopted a Heisenberg model composed of two sublattices and used the linearized spin-wave approximation around the mean-field solution. 
At $T_0 < T_c$, 1/$T_1$ increases rapidly on one sublattice, 
whereas on the other site it does not 
increase up to $T_c$, as usual in a ferromagnet. 
This is due to the fact that an $N$-type ferrimagnet has two magnon 
excitations. 
The soft magnon contributes to the increasing behavior of 
1/$T_1$ at $T_0$, since the bandwidth of the soft magnon is less than $T_0$ in 
energy. 

At low temperatures, the $T$-dependence of 1/$T_1$ is well fitted by $T^2$, 
similar to the ferromagnetic case. With increasing temperature, a $T^5$-component 
is added, due to the momentum dependence of the dispersion relation. 
In this paper, however, we have considered only the Raman process. 
When the three-magnon process, magnon-magnon interactions, and other factors are 
involved, those $T$-dependences will be modified. 
Those are beyond the purpose of this paper and will be studied 
elsewhere. 

The increase in 1/$T_1$ below $T_c$ is found also in a $P$-type 
ferrimagnet, which shows 
hump structure in the temperature dependence of the magnetization instead of  
compensation. 
Also in a $P$-type ferrimagnet, we find a rapid increase of 1/$T_1$ below 
$T_c$, even though the magnetization does not show compensation. 
This also can be explained by the fact that a $P$-type ferrimagnet is composed 
of soft and hard magnons. 
Although a $P$-type ferrimagnet does not show compensation, 1/$T_1$ on 
one sublattice still increases below $T_c$. 
We expect this to be experimentally confirmed in the near future. 

So far, we have not considered the orbital angular momentum $\vec{L}$. 
For example, in rare-earth (R) iron garnets, R$_3$Fe$_5$O$_{12}$ (R=Ho, Er, Tb, 
etc.), the  rare-earth magnetization is calculated by using the total angular momentum $\vec{J}$ $=$ $\vec{L}$ $+$ 
$\vec{S}$\cite{vanvleck32}.  
The Land{\' e} $g$-factor on an R-site is different from that on an iron site, and $T_M\neq T_A$ in general.  
Still, Eq. (\ref{heisenberg}) is our starting point.  
The expectation value of $S_A^z$ then contains the extra factor 
($\gfac_A$-1), so that $\langle S_A^z \rangle$$=$($\gfac_A$-1)$\langle J_A^z \rangle$ 
\cite{vanvleck32}, where $J_\nu^z$ is the $z$-component of $\vec{J}$ on $\nu$-site ($\nu$=A, B).
These factors can be renormalized into $J_A$, $J_B$, and $J_C$: 
$K_A\equiv (\gfac_A-1)^2 J_A$, $K_B\equiv (\gfac_B-1)^2 J_B$, and $K_C\equiv (\gfac_A-1)(\gfac_B-1) J_C$\cite{degennes58,szytula89}.
Using $K_A$, $K_B$, and $K_C$, the magnon dispersion relations are obtained by substituting $\langle J_A^z \rangle$ and $\langle J_B^z \rangle$ for $M_A$ and $M_B$, respectively. 
See also Appendix\ref{D}.
Around which temperature, $T_M$ or $T_A$, does 1/$T_1$ start to increase? 
The magnon bandwidth is determined by the expectation value of 
$\langle S_A^z\rangle=(\gfac_A-1)\langle J_A^z\rangle$ and 
$\langle S_B^z\rangle=(\gfac_B-1)\langle J_B^z\rangle$ instead of 
$\langle J_A^z\rangle$ and 
$\langle J_B^z\rangle$.
We recall that 1/$T_1$ increases, when $k_B T$ is comparable to the bandwidth, and  
$T_A$ is determined by $\langle J_A^z\rangle$ and $\langle J_B^z\rangle$.  
For example, in a case with $\gfac_A$=5/4, $\gfac_B$=2, such as for Ho$_3$Fe$_5$O$_{12}$, 
the factors $(\gfac_A-1)$ and $(\gfac_B-1)$ are smaller than 1. 
A rough estimate of the energy scale of the bandwidth \tblu{is thus} smaller than $T_A$.  
This means that 1/$T_1$ will start to increase further blow $T_A$. 
However, those energy scales are different depending on the materials \tblu{involved}. 
\tblu{Thus,} it is difficult to identify the temperature at which 1/$T_1$ starts to increase. 
Such a material dependence will be discussed in the near future and will be clarified experimentally.  
 
On the other hand, it is clear that 
$E_g$ becomes zero at $T_A$ instead of $T_M$
\begin{equation}
E_g=12 K_C \left[\langle J_A^z\rangle - \langle J_B^z \rangle \right].
\end{equation} 
Magnon excitations in RIGs \tblu{have been} reported \tblu{from} inelastic neutron scattering~\cite{plant77,nambu20}. 
However, $E_g$ \tblu{has not yet been clarified }around the compensation temperature.  
The loss of $E_g$ must be associated with the increase of domain wall speed at $T_A$\cite{jiang06,stanciu07,kjkim17,skkim19} 
and the enhancement of NMR signals\cite{imai20}. 
Such remarkable changes of the domain walls will make ferrimagnets more useful for spintronics.  
A consistent understanding \tblu{of the} NMR, ESR, and neutron-scattering \tblu{results will be even more important and }useful.  

\begin{acknowledgment}
The author thanks S. Maekawa, H. Chudo, M. Imai, M. Fujita, Y. Kawamoto, \tblu{S. Kambe, Y. Tokunaga, and H. Sakai} for 
useful and helpful discussions. 
This work was supported by Grants-in-Aid for Scientific Research (Grant 
18H04492 and 20K03810) from JSPS and MEXT, and by the inter-university 
cooperative research program of IMR Tohoku University (20N0006). A part of the 
numerical calculation was done with the supercomputer of JAEA. 
\end{acknowledgment}

\appendix
\section{Mean-field Equation}\label{B}
The mean-field equation and its solution are straightforward. 
\tblu{We define} 
\begin{align}
M_A & = f_{S_A}\left[ {\left( {{z_A}{J_A}{M_A} - zJ_C{M_B}} \right)/(k_{B}T)} 
\right],\label{mfeq1}\\
M_B & = f_{S_B}\left[ {\left( {{z_B}{J_B}{M_B} - zJ_C{M_A}} \right)/(k_{B}T)} 
\right],\label{mfeq2}\\
f_S[x] &\equiv (S+1/2)\coth\left[x(S+1/2)\right]-1/2\coth\left(x/2\right). 
\end{align} 
Note that $g$ and $\mu_B$ \tblu{have been omitted to make} the equations \tblu{clearer}. 
\tblu{Solving} Eqs. (\ref{mfeq1}) and (\ref{mfeq2}) \tblu{gives} $T_c$ \tblu{in the form}
\begin{align}
{T_c} &= \frac{1}{2}\biggl[ X_A z_A J_A + X_B z_B J_B \nonumber\\
& +  \sqrt{\left( X_A z_A J_A - X_B z_B J_B \right)^2 + X_A X_B \left( 2zJ_C 
	\right)^2 } \biggr],\\
{X_A} & \equiv (S_A + 1)S_A/3,\\
{X_B} & \equiv (S_B + 1)S_B/3,
\end{align}
For the case shown in Fig. \ref{mfsol}, $T_0$ \tblu{can be approximated} by the 
Curie temperature of the system limited to the $A$-sublattice, it is given by \tblu{$T_0\sim$} $z_A 
J_A X_A$. 

\section{\tblu{Magnon }Dispersion Relations in a Ferrimagnet}\label{A}
\begin{figure}[htbp]
	\centering
	\includegraphics[width=0.40\textwidth]{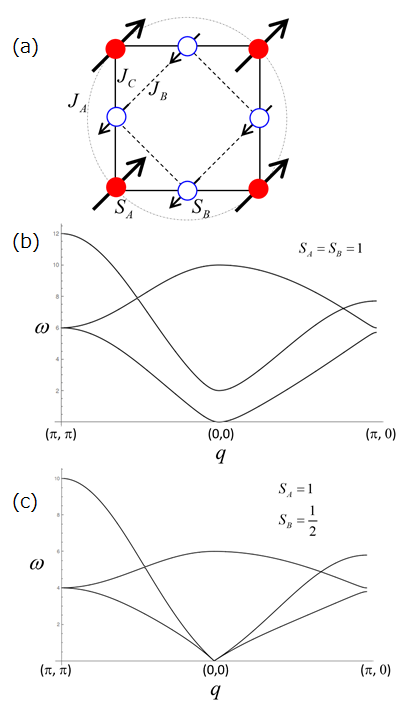}\label{lattice2}
	\caption{(Color online) (a) Schematic \tblu{illustration} of the  compensation. The 
			$A$-sublattice \tblu{has one site}, whereas the  $B$-sublattice \tblu{two sites}. 
			(b) 
			\tblu{Magnon d}ispersion relation away from the \tblu{compensation point} for $J_A$=$J_B$=$J_C$=1, $S_A$=$S_B$=1. 
			(c) \tblu{Magnon d}ispersion relation at the \tblu{compensation point} for 
			$J_A$=$J_B$=$J_C$=1, $S_A$=1 and $S_B$=1/2. In both (b) and (c), the 
			parameters \tblu{were} chosen by hand to show the characteristics. }  
\end{figure}

At low energies away from the compensation temperature, the \tblu{magnon} dispersion relation\tblu{s given by} 
Eqs. (\ref{ea}) and (\ref{eb}) \tblu{can be} expanded as
\begin{align}
E_{\alpha \bmq}
&\sim C Q^2,\\
E_{\beta \bmq}
&\sim 12J_C\left( {{M_A} - {M_B}} \right) + D{Q^2},\\
C	&=\frac{{\left[ {3{J_A}M_A^2 - \left( {{J_A} + {J_B} - 2J_C} 
\right){M_A}{M_B} + 3{J_B}M_B^2} \right]}}{{{M_A} - {M_B}}},\\
D	&=\frac{J_A M_A^2 - \left( 3J_A + 3J_B + 2J_C \right)M_A M_B + J_B M_B^2}{M_A 
- M_B},\\
\varepsilon_{1\bmq} + \varepsilon_{2\bmq} 
&= 6J_C(M_A+M_B)+4(J_AM_A+J_BM_B)Q^2,\\
\Delta_q &= 6J_C\left( M_A - M_B \right)\nonumber\\
&+\frac{12J_C\left( {{M_B} + {M_A}} \right)\left( {{J_A}{M_A} + {J_B}{M_B}} 
\right) + 24{J_C^2}{M_A}{M_B}}{  6J_C\left( M_A - M_B \right)}Q^2,\label{delta1}
\end{align}
where $Q$ $\equiv$ $\sqrt{q_x^2+q_y^2+q_z^2}$. 
Note that (0,0,0) and $(\pi,\pi,\pi)$ are equivalent \tblu{for} the case \tblu{shown in} Fig. 1. 
At the compensation temperature $M_A$ $=$ $M_B$ $\equiv$ $M$, the excitation 
gap vanishes
\begin{align}
E_{\alpha \bmq}
&\sim C_2 Q^2 + C_1 Q,\\
E_{\beta \bmq}
&\sim -C_2 Q^2 + C_1 Q,\\
C_1	&= 2\sqrt 3J_C\left(J_A + J_B + J_C \right)S,\\
C_2 &= 2\left( {{J_A} - {J_B}} \right)S,\\
\varepsilon_{1\bmq} + \varepsilon_{2\bmq} 
&= 12J_CS+4(J_A+J_B)SQ^2,\\
\Delta_{\bmq} &=	4S\sqrt{3 J_C\left(J_A + J_B +J_C \right)}Q. 
\end{align}
This does not depend on the lattice structure, as shown in Fig. \ref{lattice2}$\cdot$1.

\section{Nuclear Magnetic Relaxation in a Ferromagnet and an Antiferromagnet}\label{C}
\begin{figure}[htbp]
	\centering
	\includegraphics[width=0.45\textwidth]{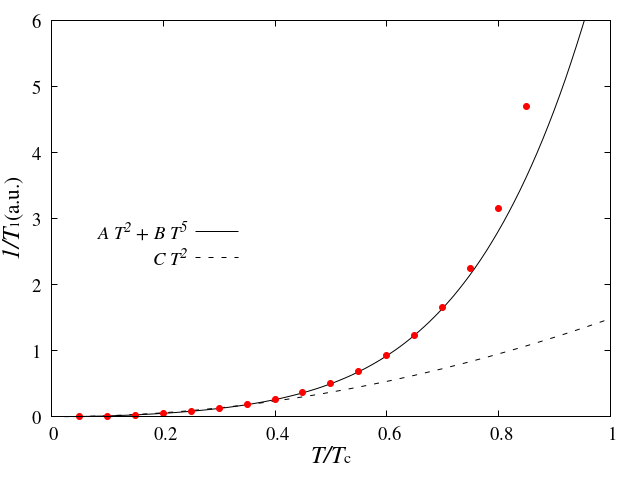}
	\caption{(Color online) The $T$-dependence of 1/$T_1$ in the ferromagnetic state. The red 
	dot\tblu{s were calculated numerically from} Eq. (\ref{t1fm}) and the mean-field 
	solution. The thick and the broken 
	lines are fitting results using $A T^2$ $+$ $B T^5$ and $C T^2$, 
	respectively. } 
\label{fmnmr}
\end{figure}

The ferromagnetic state is obtained by imposing \tblu{the conditions }$S_B$=0 and $J_C$=$J_B$=0 \tblu{on} the 
mean-field equation \tblu{given by} Eq. (\ref{heisenberg}).  
Equation (\ref{t1inv}) \tblu{then reduces to} 
\begin{align}
\frac{1}{T_{1}} 
&=2\pi F
\sum_{\bmp,\bmq}  n_B(E_{\bmp})\left[n_B(E_{\bmp})+1 \right]
\times \delta \left( E_{\bmp} - E_{\bmq} \right), \label{t1fm}
\end{align}
with $E_{\bmq}$ $=$ $z_A J_A M_A(1-\zeta_{A\bmq})$. 
The temperature dependence of 1$/T_1$ is shown in Fig. \ref{fmnmr}. 
At low temperatures, it is well fitted by $C T^2$\tblu{, because} $E_{\bmq}$ 
$\propto$ $Q^2$~\cite{beeman68,mitchell57}. On the other hand, it \tblu{deviates} 
from $T^2$ \tblu{with} increasing $T$ 
and is well 
fitted by $A T^2$ $+$ $B T^5$, since a $Q$-linear component grows in $E_{\bmq}$. 
Here, $A$, $B$, are $C$ are constants.  

\begin{figure}[htbp]
	\centering
	\includegraphics[width=0.45\textwidth]{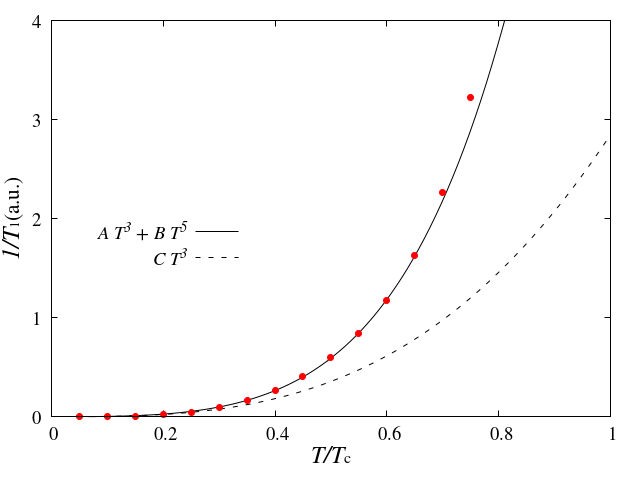}
	\caption{(Color online) The $T$-dependence of 1/$T_1$ in the \tblu{anti}ferromagnetic state. The red 
		dot\tblu{s were calculated numerically from} Eq. (\ref{t1af}) and the mean-field 
		solution. The thick and the broken 
		lines are fitting results using $A T^3$ $+$ $B T^5$ and $C T^3$, 
		respectively. } 
\label{afnmr}
\end{figure}

The antiferromagnetic state is \tblu{given} by $S_A$ = $S_B$ and $J_A$ = $J_B$ = 
0. 
In this case, $E_{\alpha \bmq}$ = $E_{\beta \bmq}$. 
Equation (\ref{t1inv}) \tblu{then reduces to}
\begin{align}
\frac{1}{T_{1,\nu}} 
&=2\pi A_\nu 
\sum_{\bmp,\bmq}  n_B(E_{\nu \bmp})\left[n_B(E_{\nu \bmp})+1 \right]
\left[ \frac{\varepsilon_{\bmp} \varepsilon_{\bmq}}{\Delta_{\bmp} \Delta_{\bmq}} + \frac{1}{4} \right]\nonumber\\
&\times \delta \left( E_{\nu \bmp} - E_{\nu \bmq} \right),\label{t1af}
\end{align}
where 
$\Delta_{\bmp}$ $=$ $2\sqrt{\varepsilon_{\bmp}^2-|\varepsilon_{3,\bmp}|^2}$
and
$\varepsilon_{\bmp}$ = $\varepsilon_{1\bmp}$ = $\varepsilon_{2\bmp}$.
Its temperature dependence is shown in Fig. \ref{afnmr}.
At low temperatures, it is well fitted by $C T^3$\tblu{, because} $E_{\bmq}$ 
$\propto$ $Q$ and  
$(\varepsilon_{\bmq}/\Delta_{\bmq})$ $\propto$ $Q^{-1}$  
~\cite{beeman68,moriya56}. 
On the other hand, it \tblu{deviates} from $T^3$ \tblu{with} increasing $T$ 
and is well fitted by $A T^3$ $+$ $B T^5$, since the curvature of $E_{\bmq}$ becomes 
relevant. 
Here, $A$, $B$, are $C$ are constants.  

\section{ESR Frequencies}\label{D}
When we consider $\vec{L}$, $J_A$, $J_B$, and $J_C$ are \tblu{replaced} by $K_A$, $K_B$, and $K_C$, and further $M_A$ and $-M_B$ are interpreted as $\langle J_A^z\rangle$ and $-\langle J_B^z\rangle$ in Eqs. (\ref{ea}) and (\ref{eb}). In a magnetic field $\vec{H}=(0,0,H)$, $\gfac_A H$ and $-\gfac_B H$ are added to Eqs. (\ref{eps1}) and (\ref{eps2}), respectively. 
The magnon excitations at $Q=0$, which correspond to the ESR frequencies $\Omega_\alpha$ and $\Omega_\beta$ are given by 
\begin{align}
\Omega_\alpha
	&=	\frac{1}{2}\left[\left(\varepsilon_{1}-\varepsilon_{2}\right)
	+\sqrt{\left(\varepsilon_{1}+\varepsilon_{2} 
		\right)^2-4\left|\varepsilon_{3}\right|^2}\right],\label{wa}\\
\Omega_\beta
	&=	\frac{1}{2}\left[-\left(\varepsilon_{1}-\varepsilon_{2}\right)
	+\sqrt{\left(\varepsilon_{1}+\varepsilon_{2} 
		\right)^2-4\left|\varepsilon_{3}\right|^2}\right].\label{wb}\\
\varepsilon_{1}
	&=  \lambda \langle J_B^z\rangle +\gfac_A H,\\
\varepsilon_{2}
	&= \lambda \langle J_A^z\rangle -\gfac_B H,\\
\varepsilon_{3}
	&= \lambda \sqrt{\langle J_A^z\rangle \langle J_B^z\rangle},
\end{align}
with $\lambda=zK_C$. 
At low temperatures\tblu{--}below \tblu{both} $T_M$ and $T_A$\tblu{--}and \tblu{to} first order in $H$,   
Eqs. (\ref{wa}) and (\ref{wb}) for $\langle J_A^z\rangle -\langle J_B^z\rangle >0$ \tblu{can be} approximated as\cite{kaplan53,wangsness53,wangsness54,tsuya54,wangsness55,geshwind59,vanvleck61}
\begin{align}
\Omega_\alpha
	&\sim \frac{\gfac_A \langle J_A^z\rangle-\gfac_B\langle J_B^z\rangle}
	        {\langle J_A^z\rangle-\langle J_B^z\rangle} H 
	 \equiv \gamma_{\rm eff} H,\\ 
\Omega_\beta
	&\sim \lambda\left(\langle J_A^z\rangle-\langle J_B^z\rangle\right) 
	      -\frac{\gfac_B \langle J_A^z\rangle-\gfac_A\langle J_B^z\rangle)}
	            {\langle J_A^z\rangle-\langle J_B^z\rangle} H,
\end{align}
with the effective gyromagnetic ratio $\gamma_{\rm eff}$. 
At $T_A$, \tblu{note} that the two frequencies become close each other, \tblu{since }$\Omega_\alpha-\Omega_\beta =(\gfac_A+\gfac_B)H$.


\begin{thebibliography}{9}
\bibitem{neel48}
	L. N\'{e}el, Ann. Phys. (Paris) {\bf 12} 137 (1948).
\bibitem{neel63}
	L. N\'{e}el, R. Pauthenet, and B. Dreyfus, 
	in {\it Progress Low Temperature Physics} ed. C.J. Gorter (North Holland, 
	Amsterdam 1964) vol. 4 Chap. VII, p.344.
\bibitem{neel71}
	L. N\'{e}el, Science {\bf 174}, 985 (1971).
	
\bibitem{gorter53}
	E. W. Gorter and J. A. Schulkes, Phys. Rev. {\bf 90}, 487 (1953).
	
\bibitem{pauthenet54} 
	R. Pauthenet and P. Blum, 
	Compt. Rend. {\bf 239}, 33 (1954).
\bibitem{bertaut56} 
	F. Bertaut and F. Forrat 
	Compt. Rend. {\bf 242}, 382 (1956).	
\bibitem{geller57acta}
	S. Geller and M. A. Gilleo, 
	Acta Cryst. {\bf 10}, 239 (1957).
\bibitem{geller57}
	S. Geller and M. A. Gilleo, 
	J. Phys. Chem. Solids {\bf 3}, 30 (1957).
\bibitem{pauthenet58} 
	R. Pauthenet, 
	Ann. Phys. {\bf 13}, 424 (1958).
\bibitem{geller63} 
	S. Geller, H. J. Williams, R. C. Sherwood, J. P. Remeika, and G. P. 
	Espinosa, 
	Phys. Rev. {\bf 131}, 1080 (1963).
\bibitem{geller65} 
	S. Geller, J. P. Remeika, R. C. Sherwood, H. J. Williams, and G. P. 
	Espinosa, 
	Phys. Rev. {\bf 137}, A1034 (1965).

\bibitem{chang65}
	J. T. Chang, J. F. Dillon, and U. F. Gianola, 
	J. Appl. Phys. {\bf 36}, 1110 (1965).
\bibitem{chow68}
	K. Chow, W. Leonard, and R. Comstock, 
	IEEE Trans. Mag. {\bf 4}, 416 (1968).
\bibitem{nelson68}
	T. Nelson, 
	IEEE Trans. Mag. {\bf 4}, 421 (1968).

\bibitem{vanwieringen53}
	J. S. van Wieringen,
	Phys. Rev. {\bf 90}, 488 (1953).
\bibitem{mcgire55}
	T. R. McGuire, 
	Phys. Rev. {\bf 97}, 831 (1955).
\bibitem{kaplan53}
	J. Kaplan and C. Kittel,
	J. Chem. Phys. {\bf 21}, 760 (1953).
\bibitem{wangsness53}
	R. K. Wangsness, 
	Phys. Rev. {\bf 91}, 1085 (1953).
\bibitem{wangsness54}
	R. K. Wangsness, 
	Phys. Rev. {\bf 93}, 68 (1954).
\bibitem{tsuya54} 
	N. Tsuya, 
	Prog Theor Phys {\bf 12}, 1 (1954).
\bibitem{wangsness55}
	R. K. Wangsness, 
	Phys. Rev. {\bf 97}, 831 (1955).
\bibitem{geshwind59}
	S. Geschwind and L. R. Walker, 
	J. Appl. Phys. {\bf 30}, S163 (1959).
\bibitem{vanvleck61}
	J. H. Van Vleck, 
	Phys. Rev. {\bf 123}, 58 (1961).
	


\bibitem{imai18}
	M. Imai, Y. Ogata, H. Chudo, M. Ono, K. Harii, M. Matsuo, Y. Ohnuma, S. 
	Maekawa, and E. Saitoh, 
	Appl. Phys. Lett. {\bf 113}, 052402 (2018).
\bibitem{imai19}
	M. Imai, H. Chudo, M. Ono, K. Harii, M. Matsuo, Y. Ohnuma, S. Maekawa, and 
	E. Saitoh, 
	Appl. Phys. Lett. {\bf 114}, 162402 (2019).
\bibitem{imai20}
	M. Imai, H. Chudo, M. Matsuo, S. Maekawa, and E. Saitoh, 
	Phys. Rev. B {\bf 102}, 014407 (2020).
\bibitem{barnet15}
	S. J. Barnett, 
	Phys. Rev. {\bf 6}, 239 (1915).

\bibitem{jiang06} 
	X. Jiang, L. Gao, J. Z. Sun, and S. S. P. Parkin, 
	Phys. Rev. Lett. {\bf 97}, 217202 (2006).
\bibitem{stanciu07}
	C. D. Stanciu, A. Tsukamoto, A. V. Kimel, F. Hansteen, A. Kirilyuk, A. 
	Itoh, and Th. Rasing, Physical Review Letters 99, 217204 (2007).
\bibitem{kjkim17}
	K.-J. Kim, S. K. Kim, Y. Hirata, S.-H. Oh, T. Tono, D.-H. Kim, T. Okuno, W. 
	S. Ham, S. Kim, G. Go, Y. Tserkovnyak, A. Tsukamoto, T. Moriyama, K.-J. 
	Lee, and T. Ono, 
	Nat. Mater. {\bf 16}, 1187 (2017).
\bibitem{skkim19}
	S. K. Kim, K. Nakata, D. Loss, and Y. Tserkovnyak, 
	Phys. Rev. Lett. {\bf 122}, 057204 (2019).

\bibitem{moriya56}
	T. Moriya, 
	Prog. Theor. Phys. {\bf 16}, 23 (1956).
\bibitem{mitchell57}
	A. H. Mitchell, 
	J. Chem. Phys. {\bf 27} 17 (1957).
\bibitem{beeman68} 
	D. Beeman and P. Pincus, 
	Phys. Rev. {\bf 166}, 359 (1968). 
		
\bibitem{tyablikov59}
	S. V. Tyablikov, 
	Ukrain. Math. Zh. {\bf 11}, 287 (1959).
\bibitem{oguchi63}
	T. Oguchi and A. Honma, 
	J. Appl. Phys., {\bf 34}, 1153 (1963).

\bibitem{mori17} 
	M. Mori, 
	J. Phys. Soc. Jpn. {\bf 86}, 124705 (2017).

\bibitem{vanvleck32}
	J. H. van Vleck, in {\it The Theory of Electric and Magnetic 
	Susceptibilities} (Oxford University Press, New York, 1932).
\bibitem{degennes58}
	P. G. de Gennes, Comptes Rendus {\bf 247}, 1836 (1958). 
\bibitem{szytula89}
	A. Szutu{\l}a, and J. Leciejewicz, 
	in {\it Handbook on the Physics and Chemistry of Rare Earths} vol. 12, eds. 
	K. A. Gschneider, Jr. and L. Eyring (Elsevier, Amsterdam, 1989) p. 131.
	
\bibitem{plant77}
	J. S. Plant, 
	Journal of Physics C: Solid State Physics {\bf 10}, 4805 (1977).
\bibitem{nambu20}
	Y. Nambu, J. Barker, Y. Okino, T. Kikkawa, Y. Shiomi, M. Enderle, T. Weber, 
	B. Winn, M. Graves-Brook, J.M. Tranquada, T. Ziman, M. Fujita, G.E.W. Bauer,
	E. Saitoh, K. Kakurai,
	Phys. Rev. Lett. {\bf 125}, 027201 (2020).

\end{thebibliography}
\end{document}